\documentclass[letterpaper,journal,onecolumn]{IEEEtran}

\usepackage{amssymb}
\usepackage{amsmath}
\usepackage{amsfonts}
\usepackage{algorithm}
\usepackage{algpseudocode}
\usepackage{epsfig}
\usepackage{graphicx}
\usepackage{color}
\usepackage{crop}
\usepackage{subfigure}
\usepackage{citesort}
\usepackage{subeqnarray}

\begin{document}

\title{An Energy-Efficient MIMO Algorithm\\with Receive Power Constraint\\ }
\author{Andr\'{e}s Alay\'{o}n Glazunov \\ 
Division of Electromagnetic Engineering\\
School of Electrical Engineering\\
KTH Royal Institute of Technology \\
Teknikringen 33, SE-100 44 Stockholm, Sweden \\}
\maketitle

\begin{abstract}
We consider the energy-efficiency of Multiple-Input Multiple-Output (MIMO) systems with constrained received power rather than constrained transmit power. A Energy-Efficient Water-Filling (EEWF) algorithm that maximizes the ratio of the transmission rate to the total transmit power has been derived. The EEWF power allocation policy establishes a trade-off between the transmission rate and the total transmit power under the total receive power constraint. The static and the uncorrelated fast fading Rayleigh channels have been considered, where the maximization is performed on the instantaneous realization of the channel assuming perfect information at both the transmitter and the receiver with equal number of antennas. We show, based on Monte Carlo simulations that the energy-efficiency provided by the EEWF algorithm can be more than an order of magnitude greater than the energy-efficiency corresponding to capacity achieving Water-Filling (WF) algorithm. We also show that the energy-efficiency increases with both the number of antennas and the signal-to-noise ratio. The corresponding transmission rate also increases but at a slower rate than the Shannon capacity, while the corresponding total transmit power decreases with the number of antennas.
\end{abstract}


\maketitle

\section{Introduction}
Energy consumption is a key issue in the deployment of \textit{green wireless} communication networks \cite{Congzheng2011}. Currently, there is a marked increase in the number of multimedia functionalities available at the mobile terminal that require ever higher transmission rates. This, together with the large screen size of mobile terminals, directly translates into a higher power consumption. On the network side, the energy-efficiency depends, among other things, on the power transmitted at the base station. Higher transmission rates require more base stations to cope with the increase in data traffic. However, the gains resulting from heavy cell splitting tend to be severely limited by high inter-cell interference. Moreover, high CAPEX (capital expenditures) as well as high OPEX (operating expenditures) associated with high power macro nodes further limits the usefulness of such an approach \cite{Glazunov2005a}.

Within this context, the introduction of the Multiple-Input Multiple-Output (MIMO) concept has become instrumental to achieve high spectral and energy-efficiencies \cite{Telatar1999}. MIMO employs multiple antenna elements at both the transmitter and the receiver and has attracted extensive attention due to the potential manifold increase in performance it can bring to standardized wireless communication networks, such as, the UMTS/LTE/LTE-A, WIMAX and WIFI systems as well as systems beyond these. Indeed, the multiple antennas can be exploited by creating a highly effective \textit{antenna diversity} system which combats the effects of fading so as to improve the average signal-to-noise ratio (SNR) or, by \textit{spatial multiplexing} of the transmit data, i.e., the data streams are spatially separated by the multiple antennas at the receiver and the transmitter \cite{Molisch_2005_textbook}.

The energy-efficiency analysis of MIMO systems has attracted the attention of the communication research community lately. A survey on energy-efficient communications can be found in \cite{BelmegaLaDe2010}. As outlined there, two main research approaches can be identified:(1) the \textit{pragmatic approach}, which focuses on system specific features such as modulation and coding-decoding schemes as well as electronics and, (2) the \textit{information-theoretic approach}, which focuses on the maximization of channel capacity per unit cost function. The latter approach has been considered in \cite{Belmega2011} where the energy-efficiency is defined as the ratio between the transmission rate and the transmit power. They show that the optimum energy-efficiency is obtained for transmit power that tends to zero for static and fast-varying MIMO channels with informed transmitter and receiver.\footnote{For the slow-fading MIMO channel they show that a non-trivial solution exists, i.e., non-zero transmit power allocation scheme.} This result illustrates the fact that according to the provided energy-efficiency definition the maximum is achieved by transmitting at very low powers. At the same time this requires the system to support very low data rates as, for example, in sensor-networks \cite{CuiGoldBahai2004,Ishmanov2011,Huang2011}. However, as they noted in \cite{Belmega2011}, this is not a practical result for communication systems where a given data rate is required for satisfactory operation of the wireless network. Hence, a different strategy is required to obtain a non-trivial solution to the energy-efficient communication over MIMO wireless channels. This is what we investigate in this article.

The solution we have found is based on the following observations. Firstly, limiting the maximum \textit{total transmit power} while maximizing the energy-efficiency, as we have seen, is not worthwhile. Secondly, we find that the transmit power can still be constrained, though indirectly, by imposing a constraint on the \textit{total receive power}. This constraint is justified by the fact, that there is a minimum power level (or signal-to-noise ratio) required at the receiver for a satisfactory operation of the communication link. Below this minimum level, which corresponds to the receiver sensitivity, the transmit signals cannot be decoded properly which leads to unwanted performance degradation. It turns out that this minimum receive power constraint can be used to obtain the power allocation for a MIMO channel that maximizes energy-efficiency.

In this paper we adhere to the information-theoretic approach outlined in \cite{Belmega2011} and find the optimal strategy for energy-efficient transition satisfying a total receive power requirement. We specialize our analysis to the static and the fast-varying single-user MIMO channels leaving the slow varying MIMO channel to be considered elsewhere. We also assume that channel state information is available at both the transmitter and the receiver. The contributions of this paper are summarized below:
\begin{itemize}
  \item We derive a Energy-Efficient Water-Filling (EEWF) algorithm that maximizes the ratio of the transmission rate to the total transmit power under total receive power constraint. This is a novel approach and differs from the customary total transmit power constraint that gives the trivial zero power result \cite{Belmega2011}.
  \item We show that the static isotropic\footnote{An isotropic channel is defined as a channel for which the likelihood of receiving a signal is the same for all directions of the unit sphere.} and the static rank-1 MIMO channels achieve the Shannon capacity if they are power efficient in the sense of the EEWF power allocation algorithm. This is obtained under the assumption of equality of the signal-to-noise ratios used in the EEWF and Water-Filling (WF). Here we show that while the optimum transmit power goes to zero as the number of antennas increases, the transmission rate also increases but to a limit defined by the minimum total received power and the noise variance.
  \item We show that static SISO channels also achieve the Shannon capacity under the conditions given above; however, for the Rayleigh fading SISO channel this is only observed in the high signal-to-noise ratio limit.
  \item We derive an upper and a lower bound for the ratio between energy-efficiency corresponding to the EEWF and the WF power allocations as well as the ratio of the transmission rate corresponding to the EEWF and the Shannon SISO capacity under the assumption of finite inverse transmit power.
  \item We illustrate for Rayleigh fading MIMO channels, based on Monte Carlo simulations, that the energy-efficiency provided by the EEWF algorithm can be more than an order of magnitude greater than the energy-efficiency corresponding to capacity achieving Water-Filling (WF) algorithm; while the corresponding total transmit power decreases with the number of antennas.
\end{itemize}

\section{Static MIMO Channel}
The input-output relationship for a MIMO communication system with $N_{\mathrm{t}}$ transmit antennas and $N_{\mathrm{r}}$ receive antennas can be written as
\begin{equation}\label{Eq:1}
\mathbf{y}=\mathbf{H}\mathbf{x}+\mathbf{n},
\end{equation}
where $\mathbf{y} \in \mathbb{C}^{N_{\mathrm{r}} \times 1}$ is the received signal vector, $\mathbf{x} \in \mathbb{C}^{N_{\mathrm{t}} \times 1}$ is the transmit signal vector, $\mathbf{H} \in \mathbb{C}^{N_{\mathrm{r}} \times N_{\mathrm{t}}}$ is the MIMO channel matrix and $\mathbf{n} \in \mathbb{C}^{N_{\mathrm{r}} \times 1}$ is the noise (AWGN) vector with covariance matrix $\mathbf{\boldsymbol{\mathcal{R}}}_{n}=\langle \mathbf{n}{\mathbf{n}}^{\dagger} \rangle=\sigma^{2}_{0} \mathbf{I}_{N_{\mathrm{r}}}$, where $\sigma^{2}_{0}$ is the noise variance, $\langle \rangle$ denotes expectation and $()^{\dagger}$ denotes the Hermitian transpose operation.

We define the \textit{total transmit power} as
\begin{equation}\label{Eq:2}
P_{t}^{x}=\mathrm{tr}\{\mathbf{\boldsymbol{\mathcal{R}}}_{x}\},
\end{equation}
where $\mathbf{\boldsymbol{\mathcal{R}}}_{x}=\langle \mathbf{x}{\mathbf{x}}^{\dagger} \rangle$ is the covariance matrix of the transmit vector and $\mathrm{tr}(\mathcal{\textbf{A}})$ denotes trace operation which is the sum of the diagonal elements of matrix $\mathcal{\textbf{A}}$.

The \textit{total receive power}\footnote{It is worthwhile to mention here that the pathloss effects can be accounted for by multiplying the channel matrix with the corresponding factor.} is obtained by fixing the channel matrix $\mathbf{H}$ and is defined for the noiseless case as
\begin{equation}\label{Eq:3}
P_{r}^{x}=\mathrm{tr}\{\mathbf{{H}^{\dagger}{H}\boldsymbol{\mathcal{R}}}_{x}\},
\end{equation}

As shown in \cite{Telatar1999}, \eqref{Eq:1} can be transformed into a stream of $r=\min\{N_{\mathrm{r}},N_{\mathrm{t}}\}$ parallel channels as follows
\begin{equation}\label{Eq:4}
\mathbf{\tilde{y}}=\mathbf{\Lambda}^{1/2}\mathbf{\tilde{x}}+\mathbf{\tilde{n}},
\end{equation}
where $\mathbf{\tilde{y}}=\mathbf{U}^{\dagger}\mathbf{y}$, $\mathbf{\tilde{x}}=\mathbf{V}^{\dagger}\mathbf{x}$, and $\mathbf{\tilde{n}}=\mathbf{U}^{\dagger}\mathbf{n}$. The matrices $\mathbf{U} \in \mathbb{C}^{N_{\mathrm{r}} \times N_{\mathrm{r}}}$ and $\mathbf{V} \in \mathbb{C}^{N_{\mathrm{t}} \times N_{\mathrm{t}}}$ are unitary and $\mathbf{\Lambda}^{1/2} = \mathrm{diag}\{\lambda_{i}^{1/2}\}  \in \mathbb{R}^{N_{\mathrm{r}} \times N_{\mathrm{t}}}$ is non-negative and diagonal containing the singular values of the MIMO channel matrix, i.e.
\begin{equation}\label{Eq:5}
\mathbf{H}=\mathbf{U}\mathbf{\Lambda}^{1/2}\mathbf{V}^{\dagger}.
\end{equation}
Hence, the transmission rate of the MIMO channel \eqref{Eq:4} equals the transmission rate of the original channel \eqref{Eq:1} due to the unitary affine transformation and is given by
\begin{equation}\label{Eq:6}
R^{x}=\log_{2}\det(\mathbf{I}_{r}+\sigma^{-2}_{0}\mathbf{\Lambda}^{1/2} \mathbf{\boldsymbol{\mathcal{R}}}_{x} \mathbf{\Lambda}^{1/2}),
\end{equation}
We measure the energy-efficiency as the ratio between the transmission rate and the transmit power (bps/Hz/Watt)\cite{Belmega2011}. Hence, we divide \eqref{Eq:6} by \eqref{Eq:2} and obtain
\begin{equation}\label{Eq:7}
\eta^{x}=\frac{R^{x}}{P_{t}^{x}}=\frac{\log_{2}\det(\mathbf{I}_{r}+\sigma^{-2}_{0}\mathbf{\Lambda}^{1/2} \mathbf{\boldsymbol{\mathcal{R}}}_{x} \mathbf{\Lambda}^{1/2})}{\mathrm{tr}\{\mathbf{\boldsymbol{\mathcal{R}}}_{x}\}},
\end{equation}

We now formulate our energy-efficiency maximization problem with constrained receive power as follows
\begin{eqnarray}\label{Eq:8}
&&\max \frac{\log_{2}\det(\mathbf{I}_{r}+\sigma^{-2}_{0}\mathbf{\Lambda}^{1/2} \mathbf{\boldsymbol{\mathcal{R}}}_{x} \mathbf{\Lambda}^{1/2})}{\mathrm{tr}\{\mathbf{\boldsymbol{\mathcal{R}}}_{x}\}},\\
&&\mathrm{s.t.}
\begin{array}{l}\label{Eq:9}
\mathrm{tr}\{\mathbf{\Lambda}^{1/2} \mathbf{\boldsymbol{\mathcal{R}}}_{x} \mathbf{\Lambda}^{1/2}\} \geq P_{r},
\end{array}
\end{eqnarray}
where we have used \eqref{Eq:3} with $P^{x}_{r} \geq P_{r}$ to get the constraint, $P_{r}$ is the minimum power required at the receiver and we use the MIMO channel power normalization $\mathrm{tr}\{\mathbf{\Lambda}\}=N_{\mathrm{r}}N_{\mathrm{t}}$.

We further specialize our analysis to the case of equal number of transmit and receive antennas, i.e., $N=N_{\mathrm{r}}=N_{\mathrm{t}}$.

Recalling that the rate \eqref{Eq:6} is maximized for independent Gaussian zero-mean complex transmit signal vector $\mathbf{x}$ with diagonal transmit covariance matrix \cite{Telatar1999}
\begin{equation}\label{Eq:10}
\mathbf{\boldsymbol{\mathcal{R}}}_{x}=\mathrm{diag}\{p_{i}\},
\end{equation}
where $p_{i}\geq 0$ is the power allocated to channel $i$.

We can now recast \eqref{Eq:8} and \eqref{Eq:9} into the following form
\begin{eqnarray}\label{Eq:11}
&&\max \frac{\sum_{i=1}^N\log_{2}(1+\sigma^{-2}_{0}\lambda_{i}p_{i})}{\sum_{i=1}^N p_{i}},\\
&& \mathrm{s.t.}
\begin{array}{l}\label{Eq:12}
  \sum_{i=1}^N \lambda_{i}p_{i} = P_{r}, \\
  p_{i}\geq 0,
\end{array}
\end{eqnarray}
where the power channel normalization is $\sum_{i=1}^N \lambda_{i} = N^2$.

The optimal transmit powers $\mathbf{p}=\left(p_{1}, p_{2}, \cdots, p_{N} \right)$ can now be obtained by solving a nonlinear programming (NLP) optimization problem defined by \eqref{Eq:11} and \eqref{Eq:12}. The solution must satisfy the Karush-Kuhn-Tucker (KKT) conditions \cite{boyd2004convex}.
We first construct the cost function containing the Lagrange multipliers $\boldsymbol{\nu}=\left(\nu_{1}, \nu_{2}, \cdots, \nu_{N} \right)$ for the inequality constraints and a multiplier $\mu$ for the equality condition
\begin{eqnarray}\label{Eq:13}
\mathcal{F}(\mathbf{p},\boldsymbol{\nu},\mu)&=&\frac{\sum_{i=1}^N\log_{2}(1+\sigma^{-2}_{0}\lambda_{i}p_{i})}{\sum_{i=1}^N p_{i}}\\
&+&\mu \left(\sum_{i=1}^N \lambda_{i}p_{i}-P_{r}\right)+\sum_{i=1}^N \nu_{i}p_{i}\nonumber,
\end{eqnarray}
The KKT conditions that maximize \eqref{Eq:13} are
\begin{eqnarray}
&&\nabla_{\mathbf{p}}\mathcal{F}(\mathbf{p},\boldsymbol{\nu},\mu) = \mathbf{0}\label{Eq:14},\\
&&\mu \left(\sum_{i=1}^N \lambda_{i}p_{i}-P_{r}\right)=0 \label{Eq:15}, \\
&&\mu \geq 0 \label{Eq:16},\\
&&\nu_{i}p_{i} = 0 \label{Eq:17},\\
&&\nu_{i} \geq 0, \label{Eq:18}
\end{eqnarray}
where $\nabla_{\mathbf{p}}=(\frac{\partial}{\partial p_{1}}, \frac{\partial}{\partial p_{2}}, \cdots, \frac{\partial}{\partial p_{N}})$ is the gradient operator.

Thus, we get from \eqref{Eq:14} that
\begin{equation}\label{Eq:19}
\frac{\frac{\sigma^{-2}_{0}\lambda_{i}\sum_{i=1}^N p_{i}}{1+\sigma^{-2}_{0}\lambda_{i}p_{i}}-\sum_{i=1}^N\ln(1+\sigma^{-2}_{0}\lambda_{i}p_{i})}{\ln(2)( \sum_{i=1}^N p_{i})^2}+\mu\lambda_{i}+\nu_{i}=0.
\end{equation}
Noticing that the constraint $p_{i} \geq 0$ need not be tight we can eliminate the constants $\nu_{i}$ since they act as slack variables. Hence, from \eqref{Eq:19} and \eqref{Eq:18} we get
\begin{equation}\label{Eq:20}
\frac{1}{\ln(2)(\frac{\sigma^{2}_{0}}{\lambda_{i}}+p_{i})} \leq \eta^{x}-\mu P_{t}^{x}\lambda_{i},
\end{equation}
where $\eta^{x}$, $P_{t}^{x}$, $\lambda_{i}$ and $\sigma_{0}$ have been defined above $i={1,,N}$. If $\eta^{x}-\mu P_{t}^{x}\lambda_{i} < \frac{\lambda_{i}}{\ln(2)\sigma^{2}_{0}}$ then \eqref{Eq:20} holds if $p_{i}>0$. Hence, we get from \eqref{Eq:17}, \eqref{Eq:18} and \eqref{Eq:19} that $p_{i}=\frac{1}{\ln(2)(\eta^{x}-\mu P_{t}^{x}\lambda_{i})}-\frac{\sigma^{2}_{0}}{\lambda_{i}}$. Now, if $\eta^{x}-\mu P_{t}^{x}\lambda_{i} > \frac{\lambda_{i}}{\ln(2)\sigma^{2}_{0}}$ then $p_{i}=0$ since $p_{i}>0$ cannot satisfy \eqref{Eq:20} and \eqref{Eq:17} at the same time.
We further see that $\sum_{i=1}^N \lambda_{i}p_{i} = P_{r}$, i.e., this constraint is tight which leads to $\mu>0$ from \eqref{Eq:15}. We have proved the following Proposition.

\textbf{\textrm{Proposition 1 (Energy-Efficient Water-Filling (EEWF)):}} Consider the energy-efficiency of MIMO communication systems operating over a static AWGN channel and constrained minimum total received power defined according to \eqref{Eq:8} and \eqref{Eq:9}. Then, for informed transmitter and receiver the optimum is achieved by a "water-filling" power allocation solution $p_{i}$ with per-channel variable power levels according to
\begin{eqnarray}
  p_{i} &=& \Biggr(\frac{1}{\ln(2)(\eta^{x}-\mu P_{t}^{x}\lambda_{i})}-\frac{\sigma^{2}_{0}}{\lambda_{i}}\Biggr)^{+}, \label{Eq:21}\\
  P_{r} &=& \sum_{i=1}^{N^{x}}\Biggr(\frac{\lambda_{i}}{\ln(2)(\eta^{x}-\mu P_{t}^{x}\lambda_{i})}- \sigma^{2}_{0}\Biggr)^{+}, \label{Eq:22}\\
  \eta^{x} &=& \frac{R^{x}}{P_{t}^{x}}, \label{Eq:23}\\
  R^{x} &=&\sum_{i=1}^{N^{x}}\log_{2}(1+\sigma^{-2}_{0}\lambda_{i}p_{i}), \label{Eq:24}\\
  P_{t}^{x} &=& \sum_{i=1}^{N^{x}} p_{i},\label{Eq:25}
\end{eqnarray}
where $\mu > 0$ is chosen to meet the receive power constraint \eqref{Eq:22}, $\eta^{x}$ is the maximum energy-efficiency and $R^{x}$ and $P_{t}^{x}$ are the corresponding transmission rate and transmit power, respectively. The summation is over $N^{x} \leq N$ channels since the number of actual channels satisfying \eqref{Eq:21} and \eqref{Eq:22} may be less than or equal to $N$. Here $(\mathbf{a})^{+}$ denotes $\max(0,\mathbf{a})$ and $\ln()$ denotes the natural logarithm.

As we can see, the optimization problem at hand involves finding $p_{i}$, $\eta^{x}$ and $\mu$. Finding $\mu$ from \eqref{Eq:22} leads to finding the roots of a polynomial of at most degree $N^{x}$ in $\mu$. Moreover, the solution is given by the largest positive real root as we show in Appendix A.

It is worthwhile to note that the transmission rate corresponding to the maximum energy-efficiency always satisfies $R^{x} \leq C$, where $C$ is the channel capacity of the static channel achieved by the water-filling (WF) algorithm, \cite{Telatar1999}. In order to be able to compare different power allocation strategies we need to define a common reference. In our case we use the transmit signal-to-noise power $SNR_{t}=\frac{\sum_{i}p_{i}}{\sigma_{0}^{2}}$ .

\textbf{\textrm{Corollary 1.1:}} The static SISO channel achieves the Shannon capacity if it is energy-efficient in the sense of Proposition 1 and it satisfies the condition of equal transmit signal-to-noise ratio such that $\frac{P_{r}}{\lambda_{1}}=P_{t}$, where, $\lambda_{1}$ is the channel gain.

The above result is straightforward. Indeed, set $N=1$ into \eqref{Eq:21}-\eqref{Eq:25} and compare it with the capacity of the AWGN SISO channel \cite{Telatar1999} for the same transmit signal-to-noise ratio $SNR_{t}=\frac{P_{t}^{x}}{\sigma^{2}_{0}}$. We see then that the EEWF power allocation policy(clearly there is only one channel) gives $p_{1}=P_{r}/\lambda_{1}$. Hence, $P_{t}^{x}=P_{r}/\lambda_{1}$, $\eta^{x} =\lambda_{1}\log_{2}(1+\frac{P_{r}}{\sigma^{2}_{0}})/P_{r}$ and $R^{x}=\log_{2}(1+\frac{P_{r}}{\sigma^{2}_{0}})$. On the other hand, the SISO channel capacity is given by $C=\log_{2}(1+\frac{P_{t}\lambda_{1}}{\sigma^{2}_{0}})$ with $p_{1}=P_{t}$ and $P_{r}^{x}=P_{t}\lambda_{1}$. Hence, $\eta^{x}_{C}=\log_{2}(1+\frac{P_{t}\lambda_{1}}{\sigma^{2}_{0}})/P_{t}$. Hence, $R^{x}=C$ when $\frac{P_{r}}{\lambda_{1}}=P_{t}$.

This result has the following interpretation (static SISO case): If the transmit and receive power constraints satisfy $\frac{P_{r}}{\lambda_{1}}=P_{t}$, then an increase or decrease of the transmit power will proportionally affect the optimum power allocation into the single link. This is not true for the general MIMO case, since the optimal power allocation in the sense of Proposition 1 and that achieving the MIMO capacity are not equivalent. Next we exemplify this latter statement by specializing our results to two special cases. We also look at their asymptotic behavior.

\textbf{\textit{Example 1: The Isotropic Static MIMO Channel.}}

Under the isotropy assumption of the static channel all the eigenvalues of the MIMO channel matrix are equal. Hence, the normalization $\sum_{i=1}^N \lambda_{i} = N^2$ gives $\lambda_{i}=N$ for $i=1,\ldots,N$. By further substitution into \eqref{Eq:21}-\eqref{Eq:25}, we readily obtain the optimal transmit power allocation per channel that achieve maximum energy-efficiency and the corresponding total transmit power and transmission rate:
\begin{eqnarray}
  p_{i} &=& \frac{P_{r}}{N^2}, \label{Eq:26}\\
  \eta^{x} &=& \frac{N^2}{P_{r}}\log_{2}(1+\frac{P_{r}}{\sigma^{2}_{0}N}), \label{Eq:27}\\
  R^{x} &=& N\log_{2}(1+\frac{P_{r}}{\sigma^{2}_{0}N}), \label{Eq:28}\\
  P_{t}^{x} &=& \frac{P_{r}}{N}, \label{Eq:29}
\end{eqnarray}
Now compare the above with the capacity achieving strategy (WF) for the isotropic static MIMO channel with transmit power constraint $\sum_{i=1}^N p_{i}=P_{t}$, \cite{Telatar1999}. We have then that $p_{i} = \frac{P_{t}}{N}$, $\eta^{x} = N \log_{2}(1+\frac{P_{t}}{\sigma^{2}_{0}})/P_{t}$, $R^{x} =C= N\log_{2}(1+\frac{P_{t}}{\sigma^{2}_{0}})$, $P_{r}^{x} = N P_{t}$. Hence, we see that energy-efficient channel also achieves the Shannon capacity under certain conditions as stated in the following result.

\textbf{\textrm{Corollary 1.2:}} The isotropic static MIMO channel achieves the Shannon capacity if it is energy-efficient in the sense of Proposition 1 and it satisfies the conditions of equal transmit signal-to-noise ratio and equal number of antennas such that $\frac{P_{r}}{N}=P_{t}$.

\textbf{\textrm{Remark 1.2.1:}} A energy-efficient, in the sense of Proposition 1, isotropic static MIMO channel with a large number of antennas behaves as follows
\begin{eqnarray}
  \lim_{N \to +\infty} p_{i} &=& 0, \label{Eq:30}\\
  \lim_{N \to +\infty} \eta^{x} &=& \infty, \label{Eq:31}\\
  \lim_{N \to +\infty} R^{x} &=& \frac{P_{r}}{\ln(2)\sigma^{2}_{0}}, \label{Eq:32}\\
  \lim_{N \to +\infty} P_{t}^{x} &=& 0, \label{Eq:33}
\end{eqnarray}
As we can see, the energy-efficiency increases asymptotically to infinity while the corresponding transmit power goes to zero and the transmit rate goes to a maximum limiting value. The physical interpretation is that if we constrain the receive power to a certain level, then the energy-efficiency will increase as we increase the number of antennas while the needed transmit power will decrease accordingly.

\textbf{\textit{Example 2: The Rank-1 Static MIMO Channel.}}

Under the rank-$1$ assumption of the static channel, all the eigenvalues of the MIMO channel matrix but one are equal zero. Hence, the normalization $\sum_{i=1}^N \lambda_{i} = N^2$ gives $\lambda_{1}=N^2$ and $\lambda_{i}=0$ for $i=2,\ldots,N$. Further substitution into \eqref{Eq:21}-\eqref{Eq:25} gives the following results
\begin{eqnarray}
  p_{1} &=& \frac{P_{r}}{N^2}, \label{Eq:34}\\
  \eta^{x} &=& \frac{N^2}{P_{r}}\log_{2}(1+\frac{P_{r}}{\sigma^{2}_{0}}), \label{Eq:35}\\
  R^{x} &=& \log_{2}(1+\frac{P_{r}}{\sigma^{2}_{0}}), \label{Eq:36}\\
  P_{t}^{x} &=& \frac{P_{r}}{N^2}, \label{Eq:37}
\end{eqnarray}

\textbf{\textrm{Corollary 1.3:}} The rank-1 static MIMO channel achieves the Shannon capacity if it is energy-efficient in the sense of Proposition 1 and it satisfies the conditions of equal transmit signal-to-noise ratio and equal number of antennas such that $\frac{P_{r}}{N^2}=P_{t}$.

This result follows by noticing that the capacity achieving strategy for the rank-1 static MIMO channel with transmit power constraint $\sum_{i=1}^N p_{i}=P_{t}$  \cite{Telatar1999}, is equivalent to the transmission rate and transmit power that corresponds to the maximum energy-efficiency and satisfying conditions above. Indeed, we have for the capacity achieving strategy that $p_{1} = \frac{P_{t}}{N^2}$ and $p_{i} = 0$ for $i=2,\ldots,N$, $\eta^{x} =\log_{2}(1+\frac{N^2 P_{t}}{\sigma^{2}_{0}})/P_{t}$, $R^{x} =C= \log_{2}(1+\frac{N^2P_{t}}{\sigma^{2}_{0}})$, $P_{r}^{x} = N^2 P_{t}$.

\textbf{\textrm{Remark 1.3.1:}} A energy-efficient, in the sense of Proposition 1, rank-1 static MIMO channel with a large number of antennas behaves as follows
\begin{eqnarray}
  \lim_{N \to +\infty} p_{1} &=& 0, \label{Eq:38}\\
  \lim_{N \to +\infty} \eta^{x} &=& \infty, \label{Eq:39}\\
  \lim_{N \to +\infty} R^{x} &=& \log_{2}(1+\frac{P_{r}}{\sigma^{2}_{0}}), \label{Eq:40}\\
  \lim_{N \to +\infty} P_{t}^{x} &=& 0, \label{Eq:41}
\end{eqnarray}
As we can see, the energy-efficiency increases asymptotically to infinity at the same time as the corresponding transmit power goes to zero and the transmit rate remains constant and equal to the capacity of the SISO channel and is independent of the number of antennas.

\section{Fast Varying MIMO Channel}
We now consider a channel that undergoes rapid fading variations and is ergodic under the transmission duration. We further assume that the elements of the channel matrix $\mathbf{H}$ in \eqref{Eq:1} are now i.i.d. zero-mean complex Gaussian variables, i.e., the channel is Rayleigh fading.

We analyze the energy-efficiency of a time-varying ergodic MIMO channel as the maximization of the instantaneous energy-efficiency given by \eqref{Eq:7} with an instantaneous constraint on the total receive power. The latter basically defines a power control constraint on the link. Hence, we define the corresponding optimization problem as follows
\begin{eqnarray}\label{Eq:42}
&&\Big\langle \max \frac{ \sum_{i=1}^N\log_{2}(1+\sigma^{-2}_{0}\lambda_{i}p_{i})}{\sum_{i=1}^N p_{i}}\Big\rangle ,\\
&& \mathrm{s.t.}
\begin{array}{l}\label{Eq:43}
  \sum_{i=1}^N \lambda_{i} p_{i} = P_{r}, \\
  p_{i}\geq 0,
\end{array}
\end{eqnarray}
where the expectation $\langle \rangle$ is taken over the realization of $\lambda_{i}$ and the maximization is over $p_{i}$. The power channel normalization is $\sum_{i=1}^N \langle \lambda_{i} \rangle= N^2$.

We can now construct a cost function and KKT conditions similar to \eqref{Eq:13} and \eqref{Eq:14}-\eqref{Eq:18}, respectively. We enforce this conditions on the instantaneous realizations of $\lambda_{i}$ and $p_{i}$. Clearly, we seek find the instantaneous $p_{i}$ that maximizes the function within the expectation brackets in \eqref{Eq:42} then we obtain the expected maximum value. It should be noted that in general $\langle \max(.)\rangle \neq \max \langle (.)\rangle$.

\textbf{\textrm{Proposition 2:}} Consider the instantaneous energy-efficiency of MIMO communication systems operating over ergodic time-varying i.i.d. Rayleigh channels with informed transmitter and receiver, and with instantaneous total receive power constraint. Then, the instantaneous maximum is achieved by the energy-efficient water-filling (EEWF) power allocation solution with per-channel variable power levels according to Proposition 1. The expected optimum energy-efficiency, the corresponding transmission rate and the total transmit power are given by
\begin{equation}
  \eta^{a} = \langle \eta^{x}, R^{a} = \langle R^{x} \rangle, P_{t}^{a} = \langle P_{t}^{x}, \rangle \label{Eq:46}
\end{equation}

where $\eta^{x}$ is the maximum of the instantaneous energy-efficiency and $R^{x}$ and $P_{t}^{x}$ are the corresponding instantaneous transmission rate and transmit power, respectively. The power allocation is performed dynamically, i.e., as per channel realization, according to a given total receive power constraint.

This result is straightforward and follows from Proposition 1 above.

In addition to \eqref{Eq:46} we also consider the average of the number of transmission channels
\begin{equation}
N^{a} = \langle N^{x} \rangle, \label{Eq:47}
\end{equation}

The transmission rate corresponding to the maximum energy-efficiency always satisfies $R^{a} \leq C_{e}$, where $C_{e}$ is the ergodic channel capacity of the time-varying i.i.d. Gaussian channel \cite{SkoglundJongren2003}.

The power efficiency criteria with total receive power constraint for the SISO channel implies channel inversion. Hence, the condition for signal-to-noise ratio equality becomes $P_{t}=\langle \frac{1}{\lambda_{1}} \rangle P_{r}$ (observe that in general $\frac{1}{\langle \lambda_{1} \rangle} \neq \langle \frac{1}{\lambda_{1}} \rangle$). For the time-varying SISO channel it can be shown that $\langle \frac{1}{\lambda_{1}} \rangle\rightarrow \infty$. However, the dynamic range of realistic channels is seldom as low as null and is often only a few tens or hundreds times larger than the sample average. Moreover, a better form of power control is obtained by only compensating for fading above a certain cutoff fade depth known as truncated channel inversion \cite{Gold1997}. So we will assume that $1 \leq \langle \frac{1}{\lambda_{1}} \rangle < \infty$ for our Rayleigh fading channels.

\textbf{\textrm{Corollary 2.1:}} Consider a energy-efficient in the sense of Proposition 2 time-varying SISO channel. Then, the ratio between the transmission rate corresponding to the energy-efficient transmission $R^{a}$ and the ergodic SISO capacity $C_{e}$ satisfy the inequalities
\begin{eqnarray}\label{Eq:48}
\frac{1}{\langle\frac{1}{\lambda_{1}} \rangle} \leq \frac{R^{a}}{C_{e}} \leq 1.
\end{eqnarray}

The corresponding bounds for the ratio between the optimum SISO energy-efficiency $\eta^{a}$ and the energy-efficiency corresponding to the capacity achieving case $\eta^{a}_{C_{e}}$ are
\begin{eqnarray}\label{Eq:49}
1 \leq \frac{\eta^{a}}{\eta^{a}_{C_{e}}} \leq \langle\frac{1}{\lambda_{1}} \rangle,
\end{eqnarray}
where we have assumed equal transmit signal-to-noise ratio such that $P_{t}=\langle \frac{1}{\lambda_{1}} \rangle P_{r}$, $1 \leq \langle \frac{1}{\lambda_{1}} \rangle < \infty$ and $\langle \lambda_{1} \rangle=1$. The upper and lower bounds in \eqref{Eq:48} and \eqref{Eq:49} are achieved at the high and low signal-to-noise ratio regimes, respectively.

The derivations are given in Appendix B.

It is worthwhile to note that if we have an ideal SISO Rayleigh channel then $\langle \frac{1}{\lambda_{1}} \rangle \rightarrow \infty$ and the (average) total transmit power will therefore also increase infinitely $P^{a}_{t} \rightarrow \infty$.

\section{Numerical Examples}
In this section we present numerical computations that illustrate the dependence of the average of the optimal energy-efficiency on the signal-to-noise ratio and the number of antennas according to Proposition 2 above. We base our computations on $10000$ realizations of the entries of the MIMO channel matrix $\mathbf{H}$ generated according to the assumption of i.i.d. Rayleigh fading. Equal number of receive and transmit antennas $N$ has been assumed with channel normalization $\langle |H_{ij}|^2 \rangle= 1$. The average signal-to-noise ratio is defined as $SNR_{t}^{a}=\frac{P_{t}^{a}}{\sigma_{0}^{2}}$, where $P_{t}^{a}$ is the average transmit power \eqref{Eq:51}.

\begin{figure}[!t]
  \begin{center}
    \scalebox{0.43}{\includegraphics*[trim = 30 90 0 90]{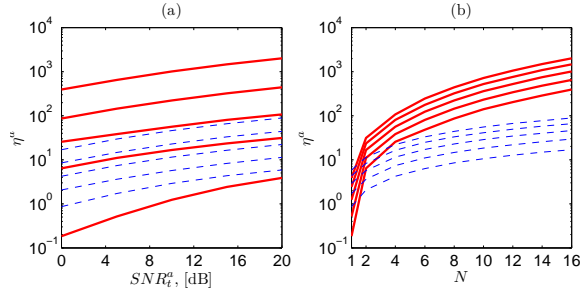}}
  \end{center}
  \caption{(a) Average energy-efficiency v.s. signal-to-noise ratio; the different curves correspond to a different number of antennas $N \in \{1,2,4,8,16\}$ going from the bottom to the top of the plot in increasing order and (b) average energy-efficiency v.s. number of antennas; the different curves correspond to different signal-to-noise ratios $SNR_{t}^{a} \in \{0,5,10,15,20\}$ dB going from the bottom to the top in increasing order. The continuous lines correspond to EEWF and the dashed lines correspond to WF.\label{F1}}
\end{figure}
\begin{figure}[!t]
  \begin{center}
    \scalebox{0.43}{\includegraphics*[trim = 30 90 0 90]{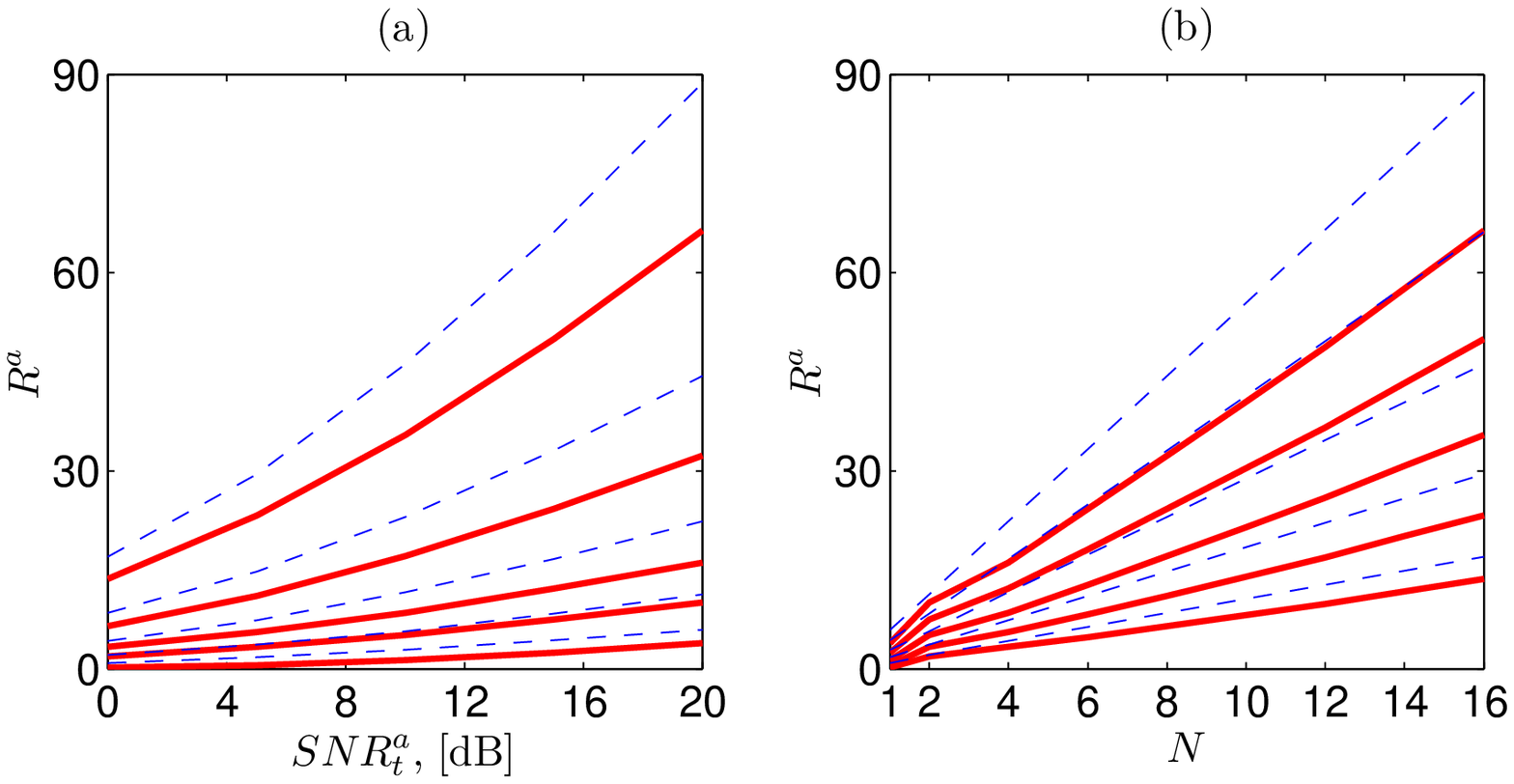}}
  \end{center}
  \caption{(a) Average transmission rate corresponding to Figure~\ref{F1} v.s. signal-to-noise ratio; the different curves correspond to a different number of antennas $N \in \{1,2,4,8,16\}$ going from the bottom to the top of the plot in increasing order and (b) average energy-efficiency v.s. number of antennas; the different curves correspond to different signal-to-noise ratios $SNR_{t}^{a} \in \{0,5,10,15,20\}$ dB going from the bottom to the top in increasing order. The continuous lines correspond to EEWF and the dashed lines correspond to WF.\label{F2}}
\end{figure}
\begin{figure}[!t]
  \begin{center}
    \scalebox{0.43}{\includegraphics*[trim = 30 90 0 90]{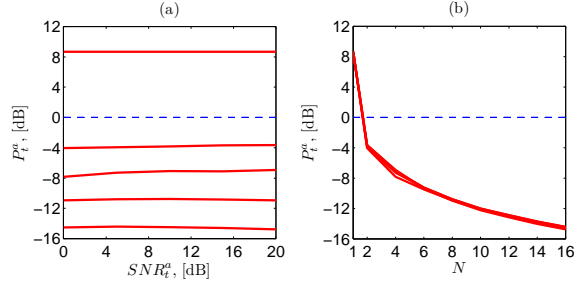}}
  \end{center}
  \caption{a) Average total transmit power corresponding to Figure~\ref{F1} v.s. signal-to-noise ratio; the different curves correspond to a different number of antennas $N \in \{1,2,4,8,16\}$ going from the top to the bottom the plot in increasing order and (b) average energy-efficiency v.s. number of antennas; the curves coincide for different signal-to-noise ratios $SNR_{t}^{a} \in \{0,5,10,15,20\}$. The continuous lines correspond to EEWF and the dashed lines correspond to WF.\label{F3}}
  \end{figure}
\begin{figure}[!t]
  \begin{center}
    \scalebox{0.43}{\includegraphics*[trim = 30 90 0 90]{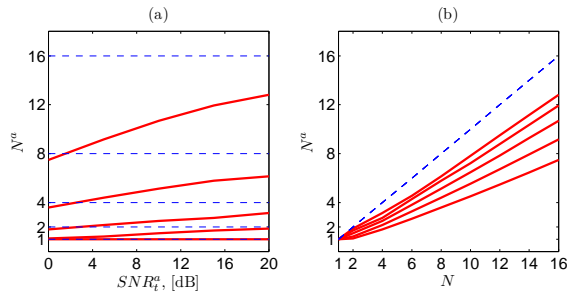}}
  \end{center}
  \caption{(a) Average number of transmission channels corresponding to Figure~\ref{F1} v.s. signal-to-noise ratio; the different curves correspond to a different number of antennas $N \in \{1,2,4,8,16\}$ going from the bottom to the top of the plot in increasing order and (b) average energy-efficiency v.s. number of antennas; the different curves correspond to different signal-to-noise ratios $SNR_{t}^{a} \in \{0,5,10,15,20\}$ dB going from the bottom to the top in increasing order. The continuous lines correspond to EEWF and the dashed lines correspond to WF.\label{F4}}
\end{figure}

Throughout the (a)-plots, from Figure~\ref{F1} to Figure~\ref{F4}, we use the following labeling. The continuous lines represent the optimum energy-efficiency solution, i.e., the EEWF algorithm with total receive power constraint $P_{r}=1$, while the dashed lines correspond to the power allocation achieving the ergodic MIMO capacity, i.e., the WF algorithm with total receive power constraint $P_{t}=1$. Furthermore, in Figure~\ref{F1}(a), Figure~\ref{F2}(a) and Figure~\ref{F4}(a) the different curves correspond to a different number of antennas $N \in \{1,2,4,8,16\}$ going from the bottom to the top of the plot in increasing order. In Figure~\ref{F3}(a) the different curves also correspond to a different number of antennas but instead the count is going from the top to the bottom of the plot in increasing order. In Figure~\ref{F1}(b), Figure~\ref{F2}(b) and Figure~\ref{F4}(b) the different curves correspond to different signal-to-noise ratios $SNR_{t}^{a} \in \{0,5,10,15,20\}$ dB also here going from the bottom to the top in increasing order.

Figure~\ref{F1}(a) and Figure~\ref{F1}(b) show the energy-efficiency $\eta^{a}$ as a function of $SNR_{t}^{a}$ and $N$, respectively. As we can see, the energy-efficiency increases as a function of both $SNR_{t}^{a}$ and $N$. Increasing the number of antennas at a fixed $SNR_{t}^{a}$ results in a larger energy-efficiency as compared to increasing the signal-to-noise ratio at a fixed $N$. Moreover, the energy-efficiency corresponding to the EEWF algorithm can be more than an order of magnitude greater than the energy-efficiency corresponding to the ergodic channel capacity. For example, in Figure~\ref{F1}(a), compare the top continuous line (EEWF) with the top dashed line (WF) for let's say $SNR_{t}^{a}=20$ dB; in both cases $N=16$. However, for the SISO case, the energy-efficiency corresponding to the WF power allocation is larger than the EEWF power allocation, which is contrary to the statement of Proposition 2 and more specifically of Corollary 2.1. Indeed, as we mentioned above, we have assumed that $P_{r}=1$ and $P_{t}=1$. Hence, they do not satisfy one of the conditions of Corollary 2.1 since $\langle \frac{1}{\lambda_{1}} \rangle \neq 1$. In this case we have that $\eta^{a}/\eta^{a}_{C_{e}}=\log_{2}(1+SNR^{a}_{t}/\langle \frac{1}{\lambda_{1}} \rangle)/\langle \log_{2}(1+\lambda_{1}SNR^{a}_{t}) \rangle \leq 1$. This follows from \eqref{Eq:65} as shown in Appendix B, where we have assumed $P_{r}=1$ for the energy-efficiency maximization and $P_{t}=1$ for the system capacity calculation. Hence, the  signal-to-noise ratio $SNR^{a}_{t}$ is given by $\langle \frac{1}{\lambda_{1}} \rangle/\sigma_{0}^{2}$ and $1/\sigma_{0}^{2}$, respectively.

For the MIMO case, i.e., $N>1$, the optimality of the EEWF power allocation comes at the expense of being suboptimal in terms of transmission rate. This is illustrated in Figure~\ref{F2}(a) and Figure~\ref{F2}(b) where the average transmission rate $R^{a}$ is plotted as a function of $SNR_{t}^{a}$ and $N$, respectively. The dashed lines denote the ergodic MIMO capacity. As expected, the transmission rate corresponding to the EEWF algorithm is bounded from above by the channel capacity. The difference increases with both the number of antennas and the signal-to-noise ratio. The latter means that the total transmit power corresponding to the EEWF must be lesser than or equal to the total transmit power corresponding to the WF. This is shown in Figure~\ref{F3}(a) and Figure~\ref{F3}(b), where the total transmit power $P_{t}^{a}$ is shown as a function of $SNR_{t}^{a}$ and $N$, respectively. On one hand, it follows from Figure~\ref{F3}(a) that the total transmit power obtained does not depend on the $SNR_{t}^{a}$. On the other hand, the total transmit power corresponding to the EEWF decreases with the number of antennas while the total receive power is fixed. Obviously, the total transmit power corresponding to WF shall not change with the number of antennas since it is a given constraint.

Figure~\ref{F4}(a) and Figure~\ref{F4}(b) show the average number of transmission channels $N^{a}$ as a function of $SNR_{t}^{a}$ and $N$, respectively. As we can see, the number of transmission channels used by the EEWF approaches asymptotically the actual number of antennas as the signal-to-noise ratio increases. However, the convergence speed decreases with the size of the MIMO arrays. The WF algorithm uses all the available channels under the i.i.d. Rayleigh fading assumption. For the EEWF algorithm this is only valid for the SISO channel.

As noted above, the maximum energy-efficiency provides a compromise between transmission rate and total transmit power. The EEWF power allocation results in reduced total transmit power at the same time as the achievable transmission rate experiments a loss with respect to the ergodic channel capacity. This trade-off should be taken into account when designing a wireless systems that aims at providing satisfactory performance in terms of both spectral- and energy-efficiency.

\section{Conclusions}
The energy-efficiency of multiple-input-multiple-output (MIMO) systems with constrained received power establishes a useful trade-off between the transmission rate and the total transmit power. The proposed energy-efficient water-filling (EEWF) algorithm provides the optimal power allocation policy that maximizes the instantaneous ratio between transmission rate and the total transmit power given a minimum required total receive power in the single-user MIMO case. Studies of the static and the uncorrelated fast fading Rayleigh channels show that the EEWF provides a non-zero optimal transmit power
and non-zero optimal transmission rate. The optimal energy-efficiency as well as the corresponding transmission rate increase with the number of antennas and the signal-to-noise ratio. On the other hand, the total transmit power decreases with the number of antennas at the same time as the receive power requirement is satisfied. Future work may include channel correlation as well as the extension to multi-user MIMO scenarios.

\section*{Appendix A}
We can, after some algebraic manipulations, arrive at the following modification of \eqref{Eq:22}
\begin{equation}\label{Eq:50}
\sum_{i=1}^{n}\frac{1}{x-\alpha_{i}}=-1,
\end{equation}
where
\begin{equation}\label{Eq:51}
\alpha_{i}=\frac{\ln(2)\eta^{x}(n\sigma^2_{0}+P_{r})}{\lambda_{i}}.
\end{equation}
The variable $x$ is related to $\mu$ as follows
\begin{equation}\label{Eq:52}
\mu=\frac{x}{\ln(2)P_{t}^{x}(n\sigma^2_{0}+P_{r})}.
\end{equation}
We can further recast \eqref{Eq:50}
\begin{equation}\label{Eq:53}
\prod_{i=1}^{n}(x-\alpha_{i})+\sum_{i=1}^{n}\prod_{j=1, j\neq i}^{n}(x-\alpha_{j})=0.
\end{equation}
Thus, solving \eqref{Eq:50} is equivalent to finding the roots of the monic polynomial of degree $n$ under the assumption that $x \neq \alpha_{i}$. Now we can expand the product terms into sums
\begin{eqnarray}\label{Eq:54}
\nonumber
P_{n}(x)&=&\sum_{k=0}^{n}a_{k}x^{k}+\sum_{i=1}^{n}\sum_{k=0}^{n-1}b_{k i}x^{k}\\
\nonumber
&=&\sum_{k=0}^{n-1}\left(a_{k}+\sum_{i=1}^{n}b_{k i}\right)x^{k}+a_{n}x^{n}\\
&=&\sum_{k=0}^{n}c_{k}x^{k}
\end{eqnarray}
where
\begin{eqnarray}
c_{n}&=& 1 \label{Eq:55}, \\
c_{k}&=&a_{k}+\sum_{i=1}^{n}b_{k i},~k=0,\ldots n-1. \label{Eq:56}
\end{eqnarray}
The coefficients $a_{k}$ are related to coefficients $\alpha_{i}$ through Vieta's formulas
\begin{eqnarray}
a_{n-1}&=&-\sum_{i=1}^{n}\alpha_{i}, \label{Eq:57} \\
a_{n-2}&=&\sum_{1\leq i < j \leq n}\!\!\!\!\!\!\alpha_{i}\alpha_{j}, \label{Eq:58}\\
\nonumber
\vdots\\
a_{n-m}&=&(-1)^{m}\!\!\!\!\!\!\!\!\!\!\!\!\sum_{\!\!\!\!\!\!1\leq i_{1} < i_{2} <\cdots< i_{m} \leq n\!\!\!}\!\!\!\!\!\!\!\!\!\!\!\!\alpha_{i_{1}}\alpha_{i_{2}}\cdots\alpha_{i_{m}}, \label{Eq:59}\\
\nonumber
\vdots\\
a_{0}&=&(-1)^{n}\prod_{i=1}^{n}\alpha_{i}, \label{Eq:60}
\end{eqnarray}
where the coefficient $a_{n-m}$ is related to a signed sum of all possible sub-products of coefficients $\alpha_{i_{m}}$, taken $m$-at-a-time also known as elementary symmetric sums. We can apply the same relationships to obtain the coefficients $b_{k i}$ from the coefficients $\alpha_{i}$ such that $i\neq k$.

Invoking the Fundamental Theorem of Algebra we have that there are $n$ complex roots that solve \eqref{Eq:52}. However, we have to satisfy the condition that $\mu$ is a positive real number. Hence, we choose the solutions that are real positive $x>0$. Furthermore, if there are several candidates we need to choose $\mu$ that actually maximizes the energy-efficiency.
Let's assume that we have two solutions $\mu_{1}$ and $\mu_{2}$ with corresponding energy-efficiencies and transmit powers $\eta^{x}_{1}$, $P_{t1}^{x}$ and  $\eta^{x}_{2}$, $P_{t2}^{x}$, respectively. From the conditions of the problem we have that $P_{r1}=P_{r2}=P_{r}$. Using this equality together with \eqref{Eq:22} and after discarding the noise term since it is asummed equal in both cases we obtain
\begin{equation}\label{Eq:61}
\sum_{i=1}^N\lambda_{i}\frac{\eta^{x}_{2}-\eta^{x}_{1}-\lambda_{i}(\mu_{2} P_{t2}^{x}-\mu_{1} P_{t1}^{x})}{(\eta^{x}_{1}-\mu_{1} P_{t1}^{x}\lambda_{i})(\eta^{x}_{2}-\mu_{2} P_{t2}^{x}\lambda_{i})} =0. \\
\end{equation}
We can set the numerator to zero for each term of the sum and since $\lambda_{i}>0$, $(\eta^{x}_{1}-\mu_{1} P_{t1}^{x}\lambda_{i})>0$ and $(\eta^{x}_{2}-\mu_{2} P_{t2}^{x}\lambda_{i})>0$ we obtain
\begin{equation}\label{Eq:62}
\eta^{x}_{2}-\eta^{x}_{1}-\lambda_{i}(\mu_{2} P_{t2}^{x}-\mu_{1} P_{t1}^{x}) = 0. \\
\end{equation}
Now, if $\eta^{x}_{2}>\eta^{x}_{1}$ and $P_{t2}^{x}<P_{t1}^{x}$ according to the premises of our problem we obtain that $\mu_{2}>\mu_{1}$ must be satisfied. Hence, the largest positive real $\mu$ gives the solution to our problem if it exists.

\section*{Appendix B}
Let's first give the transmission rate to the capacity ratio
\begin{eqnarray}\label{Eq:63}
\frac{R^{a}}{C_{e}}=\frac{\log_{2}(1+\frac{P_{t}}{\langle\frac{1}{\lambda} \rangle\sigma^{2}_{0}})}{\langle \log_{2}(1+\frac{\lambda P_{t}}{\sigma^{2}_{0}}) \rangle},
\end{eqnarray}
and the corresponding energy-efficiency ratio
\begin{eqnarray}\label{Eq:64}
\frac{\eta^{a}}{\eta^{a}_{C_{e}}}=\langle\lambda \rangle\langle\frac{1}{\lambda} \rangle\frac{\log_{2}(1+\frac{P_{t}}{\langle\frac{1}{\lambda} \rangle\sigma^{2}_{0}})}{\langle \log_{2}(1+\frac{\lambda P_{t}}{\sigma^{2}_{0}}) \rangle}.
\end{eqnarray}
In order to show the right hand side of \eqref{Eq:48} and \eqref{Eq:49} we need to show that
\begin{eqnarray}\label{Eq:65}
\log_{2}(1+\frac{P_{t}}{\langle\frac{1}{\lambda} \rangle\sigma^{2}_{0}})  \leq \langle \log_{2}(1+\frac{\lambda P_{t}}{\sigma^{2}_{0}}) \rangle,
\end{eqnarray}
where
\begin{eqnarray}\label{Eq:66}
\langle \frac{1}{\lambda}\rangle^{-1}=\frac{n}{\sum_{i=1}^{n}\frac{1}{\lambda_{i}}},
\end{eqnarray}
is the geometric mean (GM) of the discrete realizations of $\lambda$, i.e., $\lambda_{i} >0$, $i=1,\ldots,n$. Let's consider the harmonic mean-geometric mean (HM-GM) inequality
\begin{eqnarray}\label{Eq:67}
\frac{n}{\sum_{i=1}^{n}\frac{1}{\lambda_{i}\gamma}} \leq \prod_{i=1}^{n}(\lambda_{i}\gamma)^{\frac{1}{n}},
\end{eqnarray}
where $\gamma=\frac{P_{t}}{\sigma^{2}_{0}} >0$ is a constant.
Hence
\begin{eqnarray}
\!\!\!\!\!\!\!\!\!\!\!\!\log_{2}(1+\frac{n}{\sum_{i=1}^{n}\frac{1}{\lambda_{i}\gamma}})\!\!\!&\leq&\!\!\!\log_{2}(1+\prod_{i=1}^{n}(\lambda_{i}\gamma)^{\frac{1}{n}})\label{Eq:68},\\
\!\!\!&\leq&\!\!\! \log_{2}(\prod_{i=1}^{n}(1+\lambda_{i}\gamma)^{\frac{1}{n}}),\label{Eq:69}
\end{eqnarray}
since the logarithm function is a monotonically increasing function. The inequality in \eqref{Eq:69} follows from Mahler's inequality \cite{MathEncyc2}.
Finally, we readily arrive at
\begin{eqnarray}
\log_{2}(1+\frac{\gamma}{\langle\frac{1}{\lambda} \rangle}) &=& \log_{2}(1+\frac{\gamma}{\frac{1}{n}\sum_{i=1}^{n}\frac{1}{\lambda_{i}}})\label{Eq:70}\\
&\leq& \log_{2}\prod_{i=1}^{n}(1+x_{i}\gamma)^{\frac{1}{n}}\label{Eq:71}\\
&=& \frac{1}{n}\sum_{i=1}^{n}\log_{2}(1+\lambda_{i}\gamma)\label{Eq:72}\\
&=& \langle \log_{2}(1+\lambda \gamma) \rangle, \label{Eq:73}
\end{eqnarray}
The equality is achieved for infinitely large signal-to-noise ratio, i.e., $\frac{P_{t}}{\sigma^{2}_{0}}\rightarrow \infty$, which is obtained straightforwardly by applying the L'Hospital's Rule to \eqref{Eq:48}, where the derivative is taken with respect to $\frac{P_{t}}{\sigma^{2}_{0}}$.

To show the left hand side of \eqref{Eq:48} and \eqref{Eq:49} we need first to show that
\begin{eqnarray}\label{Eq:74}
\log_{2}(1+\frac{P_{t}}{\langle\frac{1}{\lambda} \rangle\sigma^{2}_{0}})  \geq \frac{1}{\langle\frac{1}{\lambda} \rangle}\log_{2}(1+\frac{P_{t}}{\sigma^{2}_{0}}).
\end{eqnarray}
But this is readily identified as the Bernoulli inequality, where $\frac{1}{\langle\frac{1}{\lambda} \rangle} \leq 1$.
Now, we see that since the logarithmic function is concave function the Jensen's inequality for the expected value reads as
\begin{eqnarray}\label{Eq:75}
\langle \log_{2}(1+\frac{\lambda P_{t}}{\sigma^{2}_{0}}) \rangle  \leq  \log_{2}(1+\frac{\langle \lambda \rangle P_{t}}{\sigma^{2}_{0}}).
\end{eqnarray}
Hence, combining \eqref{Eq:74} and \eqref{Eq:75} into \eqref{Eq:63} and \eqref{Eq:64} gives the sought-after result. In this case, the equality is obtained for infinitely small signal-to-noise ratio by following a procedure similar the described above.

\end{document}